\documentclass[12pt, prd, showpacs]{revtex4}
\usepackage{amssymb}
\usepackage{amsmath}

\input{tcilatex}

\begin{document}

\title{Super-Penrose process and rotating wormholes }
\author{O. B. Zaslavskii}
\affiliation{Department of Physics and Technology, Kharkov V.N. Karazin National
University, 4 Svoboda Square, Kharkov 61022, Ukraine}
\affiliation{Institute of Mathematics and Mechanics, Kazan Federal University, 18
Kremlyovskaya St., Kazan 420008, Russia}
\email{zaslav@ukr.net }

\begin{abstract}
We consider collision of particles in a wormhole near its throat. Particles
come from the opposite mouths. If the lapse function is small enough there,
the energy $E$ of debris at infinity grows unbounded, so we are faced with
the so-called super-Penrose process. This requires the existence of the
ergoregion, so a wormhole should be rotating.
\end{abstract}

\maketitle

In recent years, essential interest revived to high energy collisions in a
strong gravitation field. This concerns the behavior of two different
characteristics - the energy $E_{c.m.}$ in the center of mass frame of
colliding particles and/or the Killing energy $E$ of debris measured at
infinity. The first quantity can become arbitrarily large in the test
particle approximation. This was found in \cite{ban} \ (the so-called BSW
effect) for the extremal Kerr metric that provoked a huge series of works in
which the BSW effect was generalized. In spite of unbounded $E_{c.m.}$, the
quantity $E$ remains quite modest after collisions near black holes because
of strong redshift \cite{p} - \cite{fraq}.

This stimulates search for other types of objects such that $E$ (in the test
particle approximation) could be formally \ unbounded after collision (this
is called the super-Penrose process). First of all, this includes wormholes.
For the first time, high energy collisions in wormhole space-times were
considered in \cite{bambi1} where it was shown for a particular model (the
Teo wormhole \cite{teo}) that unbounded $E_{c.m.}$ is possible. It turned
out that unbounded $E$ are possible as well \cite{bambi2}. It was revealed
that there is a general underlying reasons both for unbounded $E_{c.m.}$ and 
$E$ for the Teo wormhole$.$ As is shown and extended to more general
wormhole metrics in \cite{rapid1}, \cite{rapid2}, it is connected with
extremely rapid rotation.

Recently, a work appeared in which a qualitatively new scenario is realized
in static wormholes, so rotation for getting unbounded $E_{c.m.}$ is not
required at all \cite{kras}. One particle comes from the left region, the
other one comes from the right region, so particles experience head-on
collision. Such a type of collision gives rise to unbounded $E_{c.m.}$, if
the lapse function near the throat is very small. Formally, scenarios with
head-on collisions would give unbounded $E_{c.m.}$ near black holes as well
but the problem there consists in that near the black hole horizon a
particle moves towards the horizon, not away from it, so it is difficult to
realize the head-on scenario (see Sec. IV A of \cite{fraq} for details). One
way to resolve this problem and achieve unbounded $E_{c.m.}$ due to head-on
collisions consists in considering white holes \cite{gpwhite}. The scenario
proposed by Krasnikov, enables one to find qualitatively different way to
form an initial state needed for head-on collision.

It is worth stressing that enhancement of energy requires the existence of
the ergoregion where $E$ can be negative, as usual in the Penrose process 
\cite{pen}. The conservation of energy entails that a particle with large
negative energy compensates high positive energy of debris detected at
infinity. As the Schwarzshild-like wormhole considered in \cite{kras} does
not posses the ergoregion, the class of wormholes considered by Krasnikov is
not suitable for our purposes. To achieve our goal of making the
super-Penrose process possible, we combine the Krasnikov's type of scenario
with the presence ergoregion. It can exist if rotation of a wormhole is
rapid enough. Now, we arrange head-on collision simply due to the wormhole
character of geometry only. In other realizations of the super-Penrose
process (without wormholes) it was assumed that there is a potential barrier
from which a particle can bounce back \cite{ultra}, \cite{inf}, \cite%
{ourwald}.

Let us consider the metric%
\begin{equation}
ds^{2}=-N^{2}dt^{2}+g_{\phi }(d\phi -\omega dt)^{2}+\frac{dr^{2}}{A}%
+g_{\theta }d\theta ^{2}\text{.}  \label{met}
\end{equation}%
We assume that the metric coefficient responsible for rotation $\omega >0$
like in the Kerr metric. We will consider all the processes in the plane $%
\theta =\frac{\pi }{2}$ (which is supposed to be a plane of symmetry), so
our metric effectively reduces to the 2+1 dimensional one. We also assume
that the metric coefficients do not depend on $t$ and $\phi $, so the energy 
$E=-mu_{t}$ and angular momentum $L=mu_{\phi }$ are conserved, $u^{\mu }=%
\frac{dx^{\mu }}{d\tau }$ being the four-velocity, $\tau $ the proper time.
We suppose that $r_{0}\leq r<\infty $, where $r_{0}$ is the throat radius.
In terms of the so-called shape function $b(r)$ the quantity $A=1-\frac{b}{r}
$. We make additional assumptions that $N$ has a nonzero minimum at $r=r_{0}$
and, moreover, $N_{0}=N(r_{0})\ll 1$.

The geodesic equations read%
\begin{equation}
m\frac{dt}{d\tau }=\frac{X}{N^{2}}\text{,}  \label{t}
\end{equation}%
\begin{equation}
m\frac{d\phi }{d\tau }=\frac{L}{g_{\phi }}+\frac{\omega X}{N^{2}}\text{,}
\end{equation}%
\begin{equation}
\frac{N}{\sqrt{A}}\frac{dr}{d\tau }=\sigma Z\text{,}  \label{r}
\end{equation}%
$\sigma =\pm 1$ depending on direction of motion,%
\begin{equation}
X=E-\omega L\text{, }  \label{X}
\end{equation}%
\begin{equation}
Z=\sqrt{X^{2}-N^{2}(m^{2}+\frac{L^{2}}{g_{\phi }})}\text{.}  \label{Z}
\end{equation}%
It follows from the forward-in-time condition $\frac{dt}{d\tau }>0$ that 
\begin{equation}
X>0\text{.}  \label{for}
\end{equation}

Let particles 1 and 2 come from the right and left infinities, respectively,
and collide near the wormhole throat. Then, for the energy in the centre of
mass frame we have 
\begin{equation}
E_{c.m}^{2}=-(m_{1}u_{1\mu }+m_{2}u_{2\mu })(m_{1}u_{1}^{\mu
}+m_{2}u_{2}^{\mu })=m_{1}^{2}+m_{2}^{2}+2m_{1}m_{2}\gamma \text{,}
\label{cm}
\end{equation}%
where $\gamma =-u_{1\mu }u_{2}^{\mu }$ is the Lorentz gamma factor of
relative motion.

The conservation laws read%
\begin{equation}
m_{1}u_{1}^{\mu }+m_{2}u_{2}^{\mu }=m_{3}u_{3}^{\mu }+m_{4}u_{4}^{\mu }\text{%
.}
\end{equation}%
The $t$ and $\phi $ components give us%
\begin{equation}
E_{1}+E_{2}=E_{3}+E_{4}\text{,}  \label{e}
\end{equation}%
\begin{equation}
L_{1}+L_{2}=L_{3}+L_{4}.  \label{L}
\end{equation}%
Meanwhile, the $r$ component leads to%
\begin{equation}
\sigma _{1}Z_{1}+\sigma _{2}Z_{2}=\sigma _{3}Z_{3}+\sigma Z_{4}\text{.}
\label{rz}
\end{equation}%
We assume that particle 3 escapes to the right infinity and particle 4 so
does to the left infinity. Thus $\sigma _{1}=$ $-1,\sigma _{2}=+1$, $\sigma
_{3}=+1$, $\sigma _{4}=-1$ $.$Then, the conservation of the radial momentum (%
\ref{rz}) gives us

\begin{equation}
Z_{2}-Z_{1}=Z_{3}-Z_{4}\text{.}  \label{z12}
\end{equation}

It follows from (\ref{e}) and (\ref{L}) that%
\begin{equation}
X_{1}+X_{2}=X_{3}+X_{4}\text{.}  \label{x12}
\end{equation}

Let collision happen just in the region near the throat where $N$ $\ll 1$.
We suppose that $X(r_{0})\neq 0$ and is not small for all particles. It
means that characteristics of all particles are not fine-tuned. This makes
the problem of escaping much more easy than in the black hole case (see,
e.g. \cite{j}).

Then,

\begin{equation}
Z\approx X-N^{2}z\text{, }z=\frac{1}{2X}(m^{2}+\frac{L^{2}}{g_{\phi }})\text{%
.}  \label{zn}
\end{equation}

In this region, (\ref{z12}) can be written as%
\begin{equation}
X_{2}-X_{1}\approx X_{3}-X_{4}+bN^{2}\text{,}  \label{x-2}
\end{equation}%
\begin{equation}
b=z_{4}-z_{3}+z_{2}-z_{1}\text{.}
\end{equation}

It follows from (\ref{x12}, (\ref{x-2}) that near the throat (denoted by
subscript "th")%
\begin{equation}
\left( X_{3}\right) _{th}\approx \left( X_{2}\right) _{th}-\frac{bN_{0}^{2}}{%
2},  \label{3}
\end{equation}%
\begin{equation}
\left( X_{4}\right) _{th}\approx \left( X_{1}\right) _{th}+\frac{bN_{0}^{2}}{%
2}.  \label{4}
\end{equation}

The quantities $X_{1},X_{2}$ are given by the initial conditions. Then, in
the region under discussion, one finds $X_{3}$, $X_{4}$ from (\ref{3}), (\ref%
{4}). For a given value $\left( X_{3}\right) _{th}$, there is an infinite
set of pairs $(E_{3},L_{3})$. They should obey (\ref{L}), (\ref{x12}).
Fixing, say, $L_{4}$ one finds $L_{3}$ and 
\begin{equation}
E_{3}=\left( X_{3}\right) _{th}+\omega _{th}L_{3}.  \label{e3}
\end{equation}%
Taking $L_{3}$ large, positive and unbounded, one obtains $E_{3}$ large,
positive and unbounded. This implies that, according to (\ref{e}), for fixed
finite $E_{1}$, $E_{2}$, $L_{1}$, $L_{2}$ the energy $E_{4}\rightarrow
-\infty $, $L_{4}\rightarrow -\infty $ formally. As, by assumption, there
exists the ergoregion, negative energies are admissible.

Thus we obtain simultaneously not only the analogue of the BSW effect but
also the super-Penrose process without requiring fine-tuning typical of the
BSW effect near black holes \cite{ban}. This means that rotating wormholes
can be considered as legitimate candidates for such high-energy processes.

This is not the end of story. To achieve our goal, it is necessary that
particle 3 escape to \ infinity without reflection from the potential
barrier back to the vicinity of the throat, so that the expression inside
the radical in (\ref{Z}) should be positive everywhere. Then, the condition
under discussion reads $Z^{2}>0$, where $Z$ is given by (\ref{Z}). To
simplify formulas, let us assume that $m_{3}=0$ or negligible. We have from (%
\ref{Z}) that%
\begin{equation}
X>N\frac{L}{\sqrt{g_{\phi }}}\text{,}  \label{xn}
\end{equation}%
where it is supposed that $L>0$. Eq. (\ref{xn}) is to be satisfied in an
arbitrary point, not only on the throat.

Now, we apply (\ref{xn}) to particle 3 that gives us 
\begin{equation}
X_{3}>\frac{NL_{3}}{\sqrt{g_{\phi }}}\text{.}  \label{x2}
\end{equation}%
Using (\ref{X}), we can relate them to their values on the throat:%
\begin{equation}
X_{3}=\left( X_{3}\right) _{th}+(\omega _{th}-\omega )L_{3}\text{, }
\end{equation}%
whence%
\begin{equation}
\left( X_{3}\right) _{th}+(\omega _{th}-\Omega _{+})L_{3}>0\text{,}
\end{equation}%
where by definition $\Omega _{+}=\omega +\frac{N}{\sqrt{g_{\phi }}}$. This
quantity has simple meaning. For a particle on the orbit $r=const$ the
requirement for the interval to be time-like $ds^{2}<0$, entails%
\begin{equation}
\Omega _{-}<\Omega <\Omega _{+}\text{,}  \label{range}
\end{equation}%
where in a similar way $\Omega _{-}=\omega -\frac{N}{\sqrt{g_{\phi }}},$ $%
\Omega =\frac{d\phi }{dt}$. In particular, the ergoregion appears when $%
\Omega _{-}>0$, so rotation becomes inevitable.

Discarding terms of the order $N_{0}^{2}$ and higher, we have from (\ref{3})
that $\left( X_{3}\right) _{th}\approx \left( X_{2}\right) _{th}.$
Eventually, we obtain the condition%
\begin{equation}
\left( X_{2}\right) _{th}+L_{3}(\omega _{th}-\Omega _{+})>0\text{.}
\label{om2}
\end{equation}%
As $\left( X_{2}\right) _{th}>0$ and, by assumption, $L_{3}>0$, it is
sufficient to require%
\begin{equation}
\omega _{th}-\Omega _{+}>0  \label{cond}
\end{equation}%
for (\ref{om2}) to be valid. It is worth stressing that in (\ref{cond}) the
quantity $\Omega _{+}$ is taken in the point $r$, meanwhile $\omega
_{th}=\omega (r_{0})$. Inequality (\ref{cond}) can be compatible with (\ref%
{range}). There is no contradiction here with (\ref{range}), where all
quantities are taken in the point $r$ of collision. In other words, the
sufficient condition of the validity of (\ref{om2}) reads%
\begin{equation}
\omega <\Omega _{+}<\omega _{th}\text{.}  \label{wth}
\end{equation}

For a black hole, in the horizon limit $N\rightarrow 0$ and all three
velocities $\omega $ and $\Omega _{\pm }$ tend to the same limit. By
contrast, for a wormhole case the lapse function remains small but nonzero.
Let in the vicinity of $r_{0}$,%
\begin{equation}
N^{2}\approx N_{0}^{2}+\frac{(r-r_{0})^{2}}{r_{1}^{2}}\text{,}  \label{n0}
\end{equation}%
where $r_{1}$ is some constant. Then, using the Taylor expansion in the
immediate vicinity of the throat, we have%
\begin{equation}
\Omega _{+}(r)\approx \omega _{th}+\omega ^{\prime }(r_{0})(r-r_{0})+\frac{%
N_{0}}{\sqrt{g_{\phi }(r_{0})}}.
\end{equation}

Usually, $\omega ^{\prime }<0$ (in particular, this is valid for the Kerr
and Kerr-Newman metrics). Then, we see that for a wormhole under discussion
there is a small region $r-r_{0}\leq r^{\ast }=\frac{N_{0}}{\left\vert
\omega ^{\prime }(r_{0})\right\vert \sqrt{g_{\phi }(r_{0})}}$, in which $%
0<\Omega _{+}(r)-\omega _{th}\leq \frac{N_{0},}{\sqrt{g_{\phi }(r_{0})}}$.
There, condition (\ref{cond}) can be violated. Also, eqs. (\ref{wth}) ceases
to be self-consistent there. Then, in eq. (\ref{om2}) the second term is
negative. However, the inequality can be still valid, provided $L_{3}$ is
somewhat bounded. In the worst case, for $r\rightarrow r_{0}$, it follows
from (\ref{om2}) that%
\begin{equation}
L_{3}<\left( L_{3}\right) _{\max }=\frac{\left( X_{2}\right) _{th}\sqrt{%
g_{\phi }(r_{0})}}{N_{0}}.
\end{equation}%
For a \ given $N_{0}$, this gives some upper bound. However, sending $%
N_{0}\rightarrow 0$ we obtain that $\left( L_{3}\right) _{\max }$ grows
without bound. According to (\ref{e3}), $E_{3}$ also becomes unbounded.
Thus, near the throat (where collision occurs), large $L_{3}$ is compensated
by small $N$, so $Z^{2}>0.$ Far from the throat, $X_{3}(r)$ is large due to
the term $E_{3}$ and overcomes the contribution from $L_{3}$ in (\ref{Z}),
so $Z^{2}>0$ again$.$

For small but nonzero $N_{0},$ the existence of the bound on $E_{3}$ follows
also from the Wald inequalities \cite{wald}. If, say, two massless particles
appear as a result of collision, we have for such a collisional Penrose
process (see, e.g. eq. (4) of \cite{ourwald})

\begin{equation}
2E_{3}\leq E+\sqrt{E^{2}+g_{00}E_{c.m}^{2}}\text{,}
\end{equation}%
$E=E_{1}+E_{2}$. For very large $E_{c.m}^{2}$, the maximum possible energy
at infinity%
\begin{equation}
\left( E_{3}\right) _{\max }\approx \frac{\sqrt{g_{00}}}{2}E_{c.m.}
\end{equation}

Here, $g_{00}>0$ since collision is supposed to occur in the ergoregion.

The quantities $\gamma $ and $E_{c.m.}$ (\ref{cm}) can be found directly
from equations of motion (\ref{t}) - (\ref{r}). Then, for head-on collision (%
$\sigma _{1}=-1$, $\sigma _{2}=+1$) we obtain the formula (listed in many
papers on the BSW effect)%
\begin{equation}
2m_{1}m_{2}\gamma =\frac{X_{1}X_{2}+Z_{1}Z_{2}}{N^{2}}-\frac{L_{1}L_{2}}{%
g_{\phi }}.
\end{equation}

When $N_{0}\rightarrow 0$, 
\begin{equation}
E_{c.m.}^{2}\approx \frac{2\left( X_{1}X_{2}\right) _{th}}{N_{0}^{2}}\text{.}
\end{equation}%
Taking $E_{1}=E_{2}=m\,,$ we have $E_{c.m.}\sim m/N_{0}$. Thus $\left(
E_{3}\right) _{\max }\sim m/N_{0}$ as well.

One more reservation is in order. The combination of small $N$ and a
wormhole nature of the metric (because of which the horizon is absent) leads
to the undesirable behavior of the Kretschmann invariant in the limit $%
N_{0}\rightarrow 0$. This gives a low bound on admissible value of $N_{0}$.
Corresponding estimates were made in \cite{os} for wormholes having the same
mass ($10^{5}cm$ in geometric units) as astrophysical black holes. Then, the
condition that tidal forces do not destroy atomic matter gives $%
N_{0}>10^{-13}$. This is a very weak restriction. For rotating wormholes
formulas from \cite{os} are not applicable directly but they can be used at
least for rough estimates.

Thus we showed that there is a way to achieve the super-Penrose process
without fine-tuning or specially invented scenario. Rotating wormholes
realize this in a quite natural way. It is worth stressing that the results
are model-independent and are insensitive to the details of the metric. The
key point consists in that due to a wormhole character of the metric,
head-on collision is possible \ in the region of small $N_{0}.$

\begin{acknowledgments}
This work was funded by the subsidy allocated to Kazan Federal University
for the state assignment in the sphere of scientific activities. I also
thank for support SFFR, Ukraine, Project No. 32367.
\end{acknowledgments}

\end{document}